\documentclass[journal]{IEEEtran}
\usepackage{cite}

\usepackage{graphicx}
\usepackage{amsfonts}
\usepackage{subfigure}
\usepackage{bbm}

\begin{document}

\title{When WatchDog Meets Coding\thanks{This research was supported in part by Army Research Office grant W-911-NF-0710287}}
\IEEEspecialpapernotice{(Technical Report, May 27, 2009)}

\author{\IEEEauthorblockN{Guanfeng Liang and Nitin Vaidya\\}
\IEEEauthorblockA{Department of Electrical and Computer Engineering\\
University of Illinois at Urbana-Champaign\\
Champaign, Illinois, USA\\
Email: \{gliang2, nhv\}@illinois.edu}
}


\maketitle

\begin{abstract}
In this work we study the problem of misbehavior detection in wireless networks. A commonly adopted approach is to utilize the broadcasting nature of the wireless medium and have nodes monitor their neighborhood. We call such nodes the \textit{Watchdog}s. In this paper, we first show that even if a watchdog can overhear all packet transmissions of a flow, any linear operation of the overheard packets can not eliminate miss-detection and is inefficient in terms of bandwidth. We propose a lightweigh misbehavior detection scheme which integrates the idea of watchdogs and error detection coding. We show that even if the watchdog can only observe a fraction of packets, by choosing the encoder properly, an attacker will be detected with high probability while achieving throughput arbitrarily close to optimal. Such properties reduce the incentive for the attacker to attack.
\end{abstract}

\section{Introduction}
\label{sec:intro}
In wireless ad hoc and sensor networks, paths between a source
and destination are usually multihop, and data packets are relayed in several
wireless hops from their source to their destination. This multihop nature makes the wireless networks subject to tampering attack: a compromised/misbehaving node can easily ruin data communications along the paths it is on by dropping or  corrupting packets it should forward.

Watchdog mechanism proposed in \cite{watchdog_Mobicom00} is a monitoring method used for ad hoc and sensor networks, and it is the base of many misbehavior detection algorithms and trust or reputation systems. The basic idea of watchdog is that watchdog node monitors whether its neighbor forwards the packets by overhearing. If the packet is not forwarded  within a certain period or is forward but altered, the neighbor is regarded as misbehaving in this transaction. When the misbehaving rate surpasses certain threshold, the source is notified and subsequent packets will be forwarded along other routes.

The main challenge for most watchdog mechanisms is the unreliable wireless enviorment. Due to possible reasons such as channel fading, collision with other transmission, or interference, even when the source node and the attacker are both within communication range, the watchdog may not be able to overhear every transmission and therefore is unable to determine whether there is an attack. 

To mitigate the misbehavior of the malicious nodes, a watchdog mechanism must achieve the following two goals: (1) A malicious node should be detected with high probability if it attacks. (2) The throughput under the detection mechanism should be comparable to the throughput without detection if there is no attack. These two goals seem to have conflict in interest. On one hand, to improve the probability of detection, we need to introduce more redendancy. On the other hand, better throughput requires redendancy to be reduced. 

In this paper, we show that both goals can be achieved simultaneously by introducing error detection block coding to the watchdog mechanism. This scheme is computationally simple, yet efficient. The watchdog only need to perform a compare operation. And by choosing the encoder properly, the probability of miss-detection can be made arbitrarily small while the throughput approaches optimal, even in the case when the attacker knows what encoder is being used and the watchdog can only overhear a fraction of the packets. 

The remainder of the paper is organized as follows. Section \ref{sec:related_works} discusses related work. Section \ref{sec:linear} proves any linear operation is inefficient in misbehavior detection. Section \ref{sec:single_flow} and \ref{sec:two_flows} discribe and analyse our watchdog scheme with error detection codes. 
Finally, Section \ref{sec:conclusion} concludes the paper.

\section{Related Works}
\label{sec:related_works}
To ensure the reliability of packet delivery, trust for ad hoc and sensor networks has been investigated in a lot of literatures. The foundation of such dynamic trust system is the node behavior monitoring mechanism. The most frequently used one is the watchdog mechanism proposed in \cite{watchdog_Mobicom00} and its variations.

The main idea of watchdog in \cite{watchdog_Mobicom00} was overhearing. When a node sends a packet to its neighbor, it also cached it locally. Then the node listens to its neighbor's communication. If the neighbor does not forward the same packet to its next-hop node within a short period, it is regarded as misbehaving. By this way, a node can record the successful and failed forwarding history of its next-hop.

On the basis of watchdog, various misbehavior judging and handling mechanisms are proposed. \cite{watchdog_Mobicom00} judges a node to be misbehaving when failure tally exceeds a certain threshold and it  sends a packet backward to notify the source. Then the source would choose a new route free of misbehaving node with the aid of ``pathrater''.

\cite{MNA05} proposes to measure the next-hop's behavior with the local evaluation record which is defined as a 2-tuple: packet ratio and byte ratio, forwarded by the next-hop neighbor. Local evaluation records are broadcast to all neighbors. The trust level of a node is the combination of its local observation and the broadcasted information. Trust level is inserted to the RREQ. Route is selected in the similar way to AODV \cite{AODV}. Although many ad hoc trust or reputation systems \cite{CONFIDANT_Mobihoc02}, \cite{SASN05} and \cite{Reputation_SASN04} adopt different trust level calculation mechanism, the basic processes are similar to \cite{MNA05}, including monitoring, broadcasting local observation, combing the direct and indirect information into the final trust level.

Recently, the security issue in network coding systems has drawn much attention. Due to the \textit{mixing} nature of network coding, such systems are subjects to a severe security threat, known as a \textit{pollution attack}, where attackers inject corrupted packets into the network.

Several solutions to address pollution attacks in intra-flow coding
systems use special-crafted digital signatures \cite{Kamal06signaturesfor}, \cite{Signature_Infocom08}, \cite{Zhao07signaturesfor}, \cite{batch_ICNP06} or hash
functions \cite{Krohn04on-the-flyverification}, \cite{cooperative_security_Infocom06}, which have homomorphic properties that allow
intermediate nodes to verify the integrity of combined packets.
Non-cryptographic solutions have also been proposed
\cite{Ho04byzantinemodification}, \cite{Jaggi_infocom07}. 
\cite{Jing_WiSec09} proposes two practical schemes to address pollution attacks against network coding in wireless mesh networks without requireing complex cryptographic functions and incure little overhead. 

Most of the existing network coding scheme relies on random linear combination of data packets. And as we show in Section \ref{sec:linear}, any linear operation cannot eliminate miss-detection even if all transmissions are reliable. 

\section{Limitation of Linear Coding}
\label{sec:linear}
\begin{figure}[t]
\centering
\includegraphics{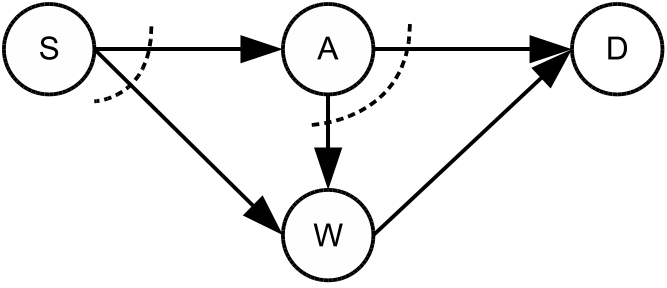}
\caption{Single Flow. Arrows (in or out) connected to the same node interfere with each other. The dash lines represent broadcast channels.}
\label{fig:single_flow}
\end{figure}
In this section, we point out the limitation for linear coding in attack detection and show the advantage of  non-linear coding. Let's consider the following example as in Fig.\ref{fig:single_flow}. There are 4 nodes in this case: the source node S, destination node D, attacker A, and the watchdog node W. Transmissions are represented by arrows. Arrows (in or out) connected to the same node interfere with each other and cannot be schedule simultaneously. The dash lines represent broadcast channels.  

Each packet consists of $n$ symbols from the finite fiele $\mathbb{F}_q$. When S (A) sends a packets, it will be received by A and W (D and W). S wants to transmit data packets to D through A. We want any tampering by A to be detected by D. 
We assume all links are reliable, have the same transmission rate 1 symbol per unit time. We also assume an optimal centralized schedule is enforced.
Under such assuptions, the watchdog W is able to monitor every packet and send $m$ checking symbols to D. The $m$ checking symbols is a funtion of $p$ and $p'$, vector representation of the original packet sent by S and the corresponding copy forwarded by A: $w = F(p,p')$. Under such assumptions the throughput is
\begin{equation}
	T = \frac{n}{2n+m} ~(symbols/unit~time).
\end{equation}

For the case of linear coding, we assume $F$ satisfies the following properties:
\begin{eqnarray}
	&F(0,0) = 0 \\
	&F(a,b) + F(c,d) = F(a+c,b+d) \\
	&F(\gamma a,0) = \gamma F(a,0)\\
	&F(0,\gamma a) = \gamma F(0,a).
\end{eqnarray}
Node D will miss an attacked packet if $F(p',p') = w$. Denote $p' = p+e$,
\begin{eqnarray}
	F(p',p') &=& F(p,p') \\
\Leftrightarrow F(p+e,p+e) &=& F(p,p+e) \\
\Leftrightarrow F(p,p) + F(e,e) &=& F(p,p) + F(0,e) \\
\Leftrightarrow F(e,e) &=& F(0,e) \\
\Leftrightarrow F(e,0) &=& 0.
\end{eqnarray}

It is easy to show that $F(e,0)$ is a linear function of $e$ and can be expressed by a $m\times n$ matrix $M$ in the finite field $\mathbb{F}_q$, and 
\begin{equation}
	F(e,0) = Me.
\end{equation}
If A chooses $e$ from the null space of M, $Null(M)$, $F(e,0)$ will be $0$ and D will consider the packet safe. Suppose A has no knowledge of $F$, the best it can do is to pick a random $e$. Then the probability of miss an attack equals to the probability of picking $e$ from the $q^{Rank(Null(M))}-1$ non-zero vectors of $Null(M)$ out of $q^n -1$ non-zero vectors in the n dimension space. Since $n-m \le Rank(Null(M))\le n$, we have the following bounds of the probability of miss-detection for any linear coding scheme
\begin{eqnarray}
	\frac{q^{n-m}-1}{2^n-1} \le &P_{miss}& \le \frac{q^n-1}{2^n-1} \nonumber\\
	\Leftrightarrow \frac{q^{n-m}-1}{q^n-1} \le &P_{miss}& \le 1.
\end{eqnarray}

So to achieve a target probability of miss-detection $\theta$, W has to send at least $m \ge \lceil-\log_2 \theta\rceil$ checking symbols to D for every packet. On other hand, if we allow $F$ to be nonlinear, only one symbol is enough to eliminate miss-detection completely. This can be easily done by setting $F(p,p') = \mathbbm{1}_{\{p=p'\}}$, which equals to 1 if $p=p'$ and 0 otherwise.

Here we want to point out the same result also applies to linear network coding. The proof is similar by considering $p$ as a \textit{generation} of $n$ coded packets and the watchdog sends $m$ linear conbinations of the packets it overhears to the receiver for verification.

\section{Single Flow Case}
\label{sec:single_flow}
Here we consider the same example in Section \ref{sec:linear}. The watchdog W will compare packets that it overhears  from both S and A, and will report an attack if they do not match. But we assume the watchdog W can detect tampering by A with probability $q$. In this case, W may not always be able to detect an attack. To enhance security, S encodes every $k$ packets into a block of $n$ coded packets with a (n,k) error detection code.  We further assume the attacker knows what encoder is being used but does not know which packets W is able to overhear.

We assume MDS (maximum distance separable) codes are being used. With a (n,k) MDS code, an attack will always be detected as long as no more than $n-k$ packets are altered. As a result, A has to alter at least $n-k+1$ packets in a block in order to avoid being detected by the decoder. And since the more packets A attacks the easier it will be caught by W, it is of A's interest to just attack the minimum number of packets per block: $n-k+1$. In this case, it is easy to show that the probobility of A not being caught is
\begin{equation}
	P_{miss}(n,k,q) = (1-q)^{n-k+1}. 
\label{eq:P_miss}
\end{equation}

We are interested in the highest coding rates we can achieve such that A has no incentive to attack. We construct a (n,k) encoder such that
\begin{equation}
	k= n+1 - \frac{f(n,q)}{q}
\label{eq:k}
\end{equation}
From Eq.\ref{eq:P_miss} we have
\begin{eqnarray}
	P_{miss}(n,k,q) &\le& e^{-q(n-k+1)} \nonumber\\
	&=& e^{-f(n,q)} 
\end{eqnarray}
We can then choose the function $f(n,q)$ appropriately so that we can make $P_{miss}$ arbitrarily small. For example, by making  $f(n,q) = \beta\ln n$ for any positive constant $\beta$, we have
\begin{eqnarray}
	P_{miss}(n,k,q) &\le& e^{-\beta \ln n} \nonumber\\
	&=& n^{-\beta} 
\label{eq:P_miss->0}
\end{eqnarray}
So we can reduce the incentive for A to attack by making the block longer. And the coding rate becomes
\begin{eqnarray}
	R &=& \frac{k}{n} \nonumber\\
	&=& \frac{n+1-\frac{\beta \ln n}{q}}{n} \nonumber \\
	&=& 1 + \frac{1}{n} - \frac{\beta}{q}\frac{\ln n}{n} 
\end{eqnarray}

\begin{figure}[t]
\centering
\includegraphics[width = 3.5 in]{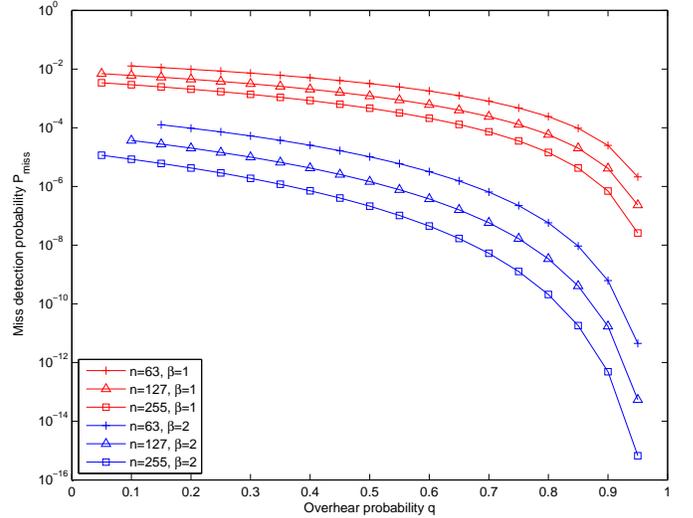}
\caption{Miss detection probability v.s. observe probability in the single flow example.}
\label{fig:plot_single_flow}
\end{figure}

\begin{figure}[t]
\centering
\includegraphics[width = 3.5 in]{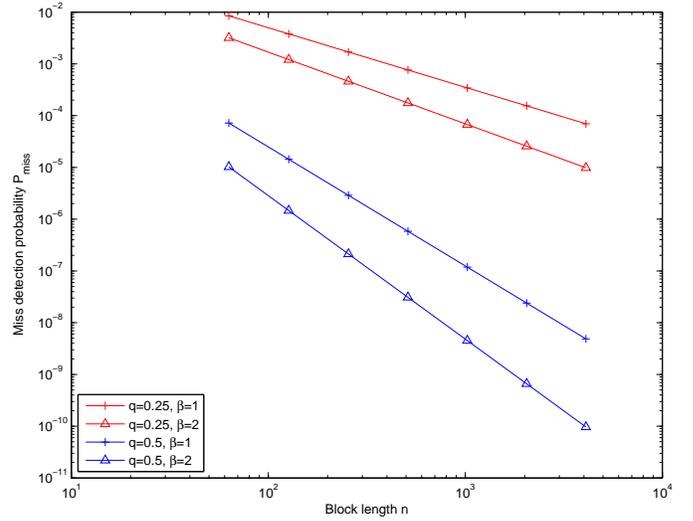}
\caption{Miss detection probability with $k = n+1-\frac{\beta \ln n}{q}$ in the single flow example.}
\label{fig:plot_single_flow2}
\end{figure}

Since the delay to verify a block equals to the time it takes to transmit $n$ packets in the block, tradeoff between probability of miss-detection and $n$ we plot in Figure \ref{fig:plot_single_flow} and Figure \ref{fig:plot_single_flow2} is also the tradeoff between miss-detection and delay. We assume
that for the n plotted in the figures, a suitable MDS (n,k) code exists for the block. We can see that by integrating a watchdog and error detection coding, we can reduce the incentive for the attacker to attack by allowing longer delay. 

Notice that by making $n$ large, the coding/decoding complexity increases. In the case complexity is a concern, the source can scramble coded packets of multiple $(n,k)$ encoded blocks and transmit these packets in a random order. By doing so, the attacker will have to corrupt more packets in order to destroy a particular block, which makes it easier to be detected by the watchdog.


\section{Two Flows Case}
\label{sec:two_flows}
\begin{figure}[t]
\centering
\includegraphics{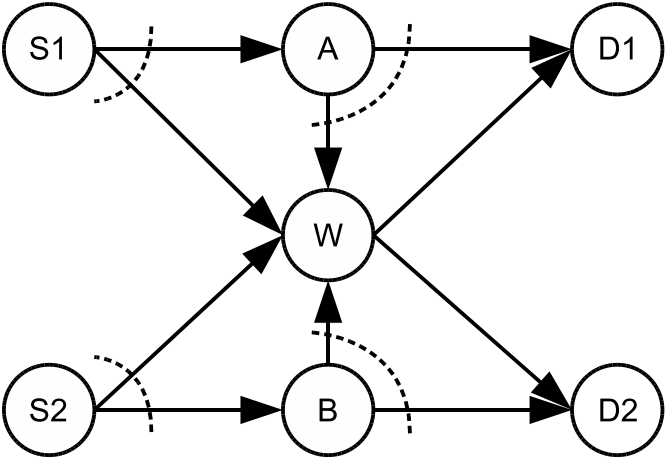}
\caption{Two Flows}
\label{fig:two_flow}
\end{figure}
In the previous section, we assume the watchdog W can only compare a packet with probability $q$. Possible reasons for making this assuption are: a watchdong node may be intentionally turned off occasionally in order to save power, or  interference from other nodes in the network makes the watchdog can observe only a fraction of the packets. In this section, we will look into the latter case. Since the level of intereference is highly correlated to the traffic load in the system, we will mainly focus on the trade-off between throughput and security.

Consider the following example. There are two flows in the system: S1-A-D1 and S2-B-D2. These flows are far away from each other so there is no inter-flow interference. But the watchdog W is sitting between the flows and can overhear transmissions on all the four links. So even though a transmission is successful along its path, it may collide with transmissions along the other flow at W. Suppose A is the attacker, we want to know the probability $q$ in this case. For traffic pattern, we assume a slotted aloha with access probability $\alpha$. To simplify the analysis, we further assume a node will access the channel by transmitting dummy packets when it has no data packet to send. Under these assumptions, we can compute the throughput and observe probability as
\begin{equation}
	T = \alpha(1-\alpha),
\label{eq:TPT}
\end{equation}
\begin{equation}
	q = (1-\alpha)^5.
\label{eq:q}
\end{equation}
The exponent in Eq.\ref{eq:q} is 5 because given that the transmission from S1 to A is successful, W can overhear it if neither S2 nor B transmit which occurs with probability $(1-\alpha)^2$. To compare this packet, W should overhear the transmission from A to D1 too, which happens with probability $(1-\alpha)^3$ for S1, S2 and B to remain siilent.

Similar to the one-flow example, we can make $P_{miss}$ arbitrarily small by choosing 
\begin{equation}
	k = n+1 - \frac{\beta \ln n}{(1-\alpha)^5}.
\label{eq:k_2flows}
\end{equation}
And the effective throughput is
\begin{eqnarray}
	T_E &=& TR \nonumber\\
&=& \alpha(1-\alpha)(1+\frac{1}{n}) - \frac{\alpha \beta \ln n}{(1-\alpha)^4 n}.
\end{eqnarray}

\begin{figure}[t]
\centering
\includegraphics[width = 3.5 in]{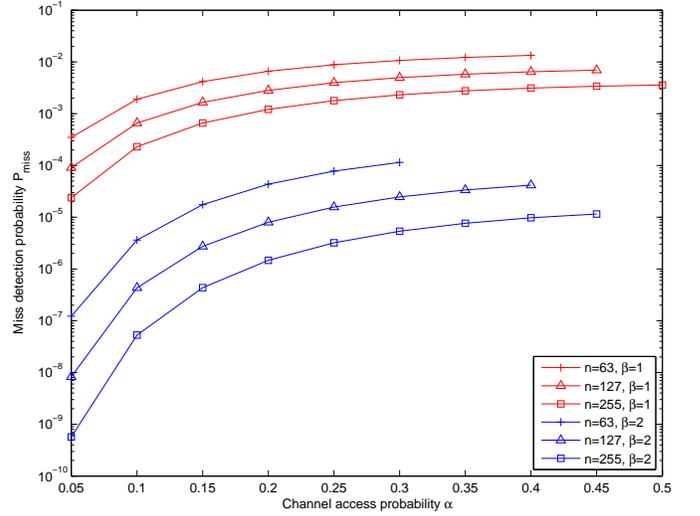}
\caption{Miss detection probability v.s. channel access probability with $k = n+1-\frac{\beta \ln n}{(1-\alpha)^5}$ in the two flows example. Where the curves stop means no code is available.}
\label{fig:plot_two_flow_P}
\end{figure}

\begin{figure}[t]
\centering
\includegraphics[width = 3.5 in]{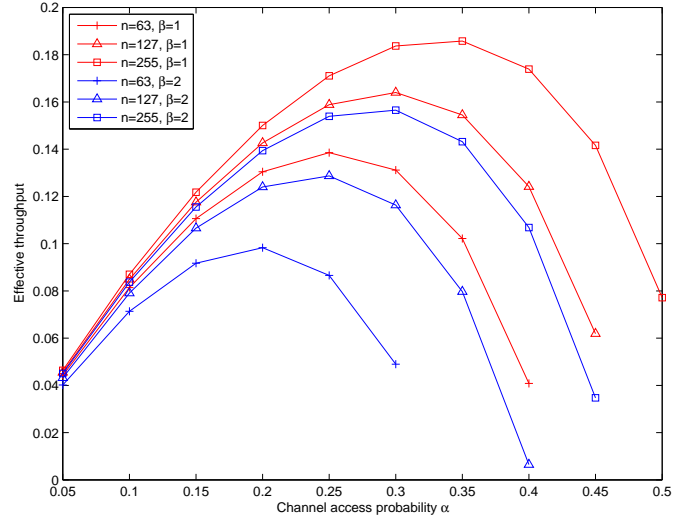}
\caption{Effective throughput v.s. channel access probability $k = n+1-\frac{\beta \ln n}{(1-\alpha)^5}$ in the two flows example. Where the curves stop means no code is available.}
\label{fig:plot_two_flow_T}
\end{figure}

In Figure \ref{fig:plot_two_flow_P} and Figure \ref{fig:plot_two_flow_T} we plot the miss-detection probability and effective throughput when the error detection code is chosen according to Eq. \ref{eq:k_2flows}. We only plot the result for $\alpha \le 0.5$ because further increasing $\alpha$ will only reduce the  throughput. We can see from Figure \ref{fig:plot_two_flow_P} the probability of miss-detection increases as the $\alpha$ increases and converges to roughly $n^{-\beta}$. Since the higher $\alpha$ is, the fewer packets the watchdog can observe, the source has to sacrify coding rate in order to maintain a certain probability of missing an attack as $\alpha$ increases. As it is shown in Figure \ref{fig:plot_two_flow_T}, as $\alpha$ increases, the effective throughput increases up to a certain level then drops to zero as $\alpha$ gets larger.


\begin{figure}[t]
\centering
\includegraphics[width = 3.5 in]{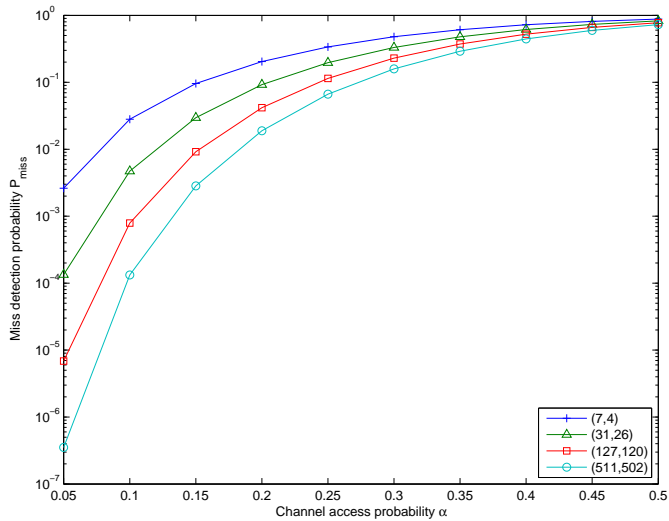}
\caption{Miss detection probability v.s. channel access probability for some Hamming codes.}
\label{fig:plot_two_flow2_P}
\end{figure}

\begin{figure}[t]
\centering
\includegraphics[width = 3.5 in]{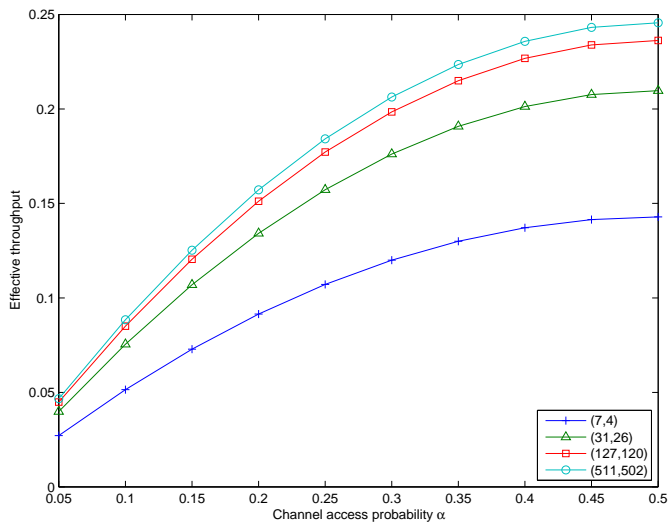}
\caption{Effective throughput v.s. channel access probability for some Hamming codes.}
\label{fig:plot_two_flow2_T}
\end{figure}

We show the performance of some $(2^m-1, 2^m-m-1)$ Hamming codes in Figure \ref{fig:plot_two_flow2_P} and Figure \ref{fig:plot_two_flow2_T}. In the case we cannot adapt the encoder to channel access probability, although there is no guarantee for miss-detection probability, a longer code always performs better in terms of both miss-detection probability and effective throughput. But such improvement comes with the cost of additional delay.

\section{Discussion}
\label{sec:more}
In the previous sections, we have studied the case when the watchdog node is trustworthy. But in reality, it is also possible that the watchdog misbehaves. We admit that our scheme may fail detecting an attack both the watchdog and the forwarder can be malicious. In this case the relay node can alter the packets as much as possible without being detected as long as the faulty watchdog never declares an attack. However, in the case of single failure (at most one of the two nodes - forward or watchdog - is faulty), if the  watchdog is faulty, the only way for it to attack the system is to accuse the relay node of attacking; and if the watchdog is well-behaving, it will declare an attack if and only if the relay node alters the packets. So under the assumption of single failure, we can be sure that either the watchdog or the relay is malicious. However, our scheme still cannot determine which node is misbehaving. To break the tie, the relay may have to be monitored by more than one watchdog and have a higher connectivity requirement. This is one of the potential directions, and we are currently working on it.

\section{Conclusion}
\label{sec:conclusion}
In this work we study the problem of misbehavior detection in wireless networks. We first show that even if a watchdog can overhear all packet transmissions of a flow, any linear operation of the overheard packets can not eliminate miss-detection and is inefficient in terms of bandwidth. We propose a lightweigh misbehavior detection scheme which integrates the idea of watchdogs and error detection coding. We show that even if the watchdog can only observe a fraction of packets, by choosing the encoder properly, an attacker will be detected with high probability while achieving throughput arbitrarily close to optimal.


\bibliographystyle{IEEEtran}
\bibliography{PaperList}
\end{document}